\documentclass{PoS}

\usepackage{amssymb,amsmath,amsfonts,amsthm,mathrsfs}
\usepackage{array,multirow}

\usepackage{graphicx}
\usepackage{epstopdf} 
\graphicspath{{./}}

\usepackage{wrapfig}

\newcommand{\Fig}[1]{Fig.~\ref{#1}}

\newcommand{\Eq}[1]{Eq.\,\eqref{#1}}

\newcommand{\Tr}{\operatorname{Tr}}

\title{
Entanglement entropy of SU(3)  Yang-Mills theory
}

\ShortTitle{Entanglement entropy}

\author{Y.~Nakagawa
\\
Research Institute for Information Science and Education,
Hiroshima University, Higashi-Hiroshima, Hiroshima, 739-8521, Japan   
\\
E-mail: \email{nkgw@rcnp.osaka-u.ac.jp}
}

\author{A.~Nakamura
\\
Research Institute for Information Science and Education,
Hiroshima University, Higashi-Hiroshima, Hiroshima, 739-8521, Japan
\\
E-mail: \email{nakamura@riise.hiroshima-u.ac.jp}
}
\author{S.~Motoki
\\
GraduateSchoolofBio-SphereScience, HiroshimaUniversity, 1-7-1Kagamiyama, 
Higashi-Hiroshima739-8521, Japan
\\
E-mail: \email{motoki-shinji@hiroshima-u.ac.jp}
}
\author{V.I. Zakharov
\\
ITEP, B. Cheremushkinskaya 25, Moscow, 117218, Russia \\
Max-Planck-Institut f\"ur Physik, F\"ohringer Ring 6, 80805 M\"unich, Germany
\\
E-mail: \email{xxz@mppmu.mpg.de}
}

\abstract{
We calculate the entanglement entropy
using a SU(3) quenched lattice gauge simulation.
We find that the entanglement entropy scales as $1/l^2$
at small $l$ as in the conformal field theory.
Here $l$ is the size of the system, whose degrees of
freedom is left after the other part are traced out.
The derivative of the entanglement entropy with respect to $l$
hits zero at about $l^{\ast} = 0.6 \sim 0.7$ [fm] and vanishes
above the length.
It may imply that the Yang-Mills theory has the mass gap
of the order of $1/l^{\ast}$.
Within our statistical errors, no discontinuous change
can be seen in the entanglement entropy.

We discuss also a subtle point appearing in gauge systems
when we divide a system with cuts.
}

\FullConference{The XXVII International Symposium on Lattice Field Theory - LAT2009\\
		 July 26-31 2009\\
		 Peking University, Beijing, China}

\begin{document}

\section{Introduction}
\label{sec:Introduction}
\vspace{-0.3cm}

Entanglement properties of quantum systems have been received much attention
in quantum information theory and condensed matter physics.
A simple system composed of two spin-1/2 system in spin singlet state
is a typical example of the entangled state.
Entanglement entropy is one of quantities measuring quantum entanglement.
It can be defined in any quantum systems, including quantum mechanical systems
and quantum field theories.
The entanglement entropy (also called geometric entropy) between two regions,
a subregion $A$ of size $l$ and its complement $B$,
measures how much two regions are quantumly correlated
and it is expected to be very useful to investigate phase structures of quantum system.

Quantum entanglement of ground states has been widely studied
in condensed matter physics
(for a review, see \cite{Amico:2007ag}).
In the Ising chain model, for instance, the entanglement entropy
at the critical point diverges while it saturates in the non-critical regime.
The entanglement entropy can serve as an order parameter
of quantum phase transitions.

As stated above, entanglement entropy can be defined in quantum field theories.
The pure Yang-Mills theory is particularly interesting
since it is a confining theory and is expected to have a mass gap.
Recently, gauge/gravity duality has been extensively studied and
it provides a method to study non-perturbative infrared dynamics
of confining gauge theories.
The calculation of the entanglement entropy using holographic
approach has been proposed by Ryu and Takayanagi
\cite{Ryu:2006rm}
(for a review on the holographic calculation. see
\cite{Nishioka:2009rz}
), and generalized by Klebanov et al.\cite{Klebanov:2008tg},
as the minimal surface $\gamma$ bending down to the bulk space,
\begin{equation}
S_A = \frac{1}{4G_N^{10}} \int_{\gamma} d^d\sigma
e^{-2\phi} \sqrt{G_{\textrm{ind}}^{(8)}}.
\end{equation}
Here $G_N^{10}$ is the 10 dimensional Newton constant,
$G_{\textrm{ind}}^{(8)}$ the induced string frame metric on
the surface $\gamma$, and $\phi$ the dilaton field.
Although gravitational background dual to pure Yang-Mills theory
has not been discovered, some approaches have been proposed
including an effective model, so-called AdS/QCD.
The numerical simulations of the entanglement entropy will
give valuable numerical support to this kind of holographic approach.

\begin{figure}
\begin{center}
\resizebox{0.3\textwidth}{!} 
{\includegraphics{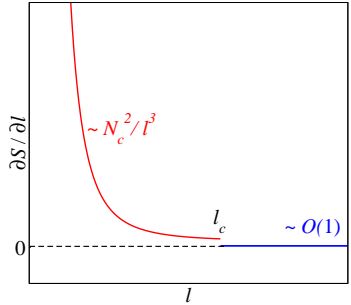}}
\vspace{-0.3cm}
\caption{
Schematic picture of the entanglement entropy
predicted by the holographic approach for the AdS bubble solution.
At small $l$, $\partial S_A/\partial l$ behaves as $1/l^3$
as conformal field theories in $(3+1)$-dimensional spacetime.
By contrast, it vanishes at large $l$ where disconnected surfaces
dominate.
}
\label{fig:dSdl_AdS_CFT}
\end{center}
\vspace{-0.5cm}
\end{figure}

In the holographic approach, the calculation of the entanglement entropy
in the gauge theory side is reduced to the calculation
of geodesics in the gravity side.
The boundary of geodesics coincides the boundary of partitioned subsystems.
This is quite similar to the calculations of the Wilson loop
in the holographic approach.
On the gravity side,
Wilson loops are obtained by the action of the string world sheet
whose boundary is the Wilson loop.
The entanglement entropy has been studied for various confining backgrounds
\cite{Klebanov:2008tg,Nishioka:2007zl}.
For some of confining backgrounds, two surfaces, called connected
and disconnected surfaces, compete and the former dominates
at small $l$.
At some critical length $l^{\ast}$, disconnected surface dominates
and the entanglement entropy becomes $l$-independent at large $l$.
The expected behavior of the entanglement entropy for the AdS bubble solution
is schematically depicted in \Fig{fig:dSdl_AdS_CFT}.
The connected surface is the order of $N_c^2$ in large $N$ expansion
while disconnected surfaces are $O(1)$.
This indicates that the effective degrees of freedom at small $l$
are gluonic degrees of freedom.
By contrast, those at large $l$ are glueballs, color singlet objects.
Therefore, it may be natural to expect that the critical length $l^{\ast}$
plays the role of (the inverse of) the critical temperature $T_c$ of
the confinement/deconfinement phase transition.

The entanglement entropy in SU(2) lattice gauge theory has been studied
by 
Velytsky \cite{Velytsky:2008rs} and
Bividovich and Polikarpov \cite{Buividovich:2008kq}.
In Ref.\cite{Velytsky:2008rs}, SU(N) lattice gauge theories are
studied in Migdal-Kadanoff approximation, 
and in Ref.\cite{Buividovich:2008kq}, SU(2) lattice gauge theory
is numerically investigated, and there is an indication
that the derivative of the entanglement entropy shows a discontinuous
change at some critical length scale $l^{\ast}$ and it vanishes.

In this paper, we investigate the entanglement entropy
in SU(3) pure Yang-Mills theory using lattice Monte Carlo simulations.
Instead of directly calculating the entropy, we adopt numerical technique
to evaluate the entanglement entropy, which has also been used in
\cite{Buividovich:2008kq}
(originally proposed in
\cite{Endrodi:2007dq,Fodor:2007sy}
in order to calculate the pressure in the deconfined phase).

\section{Definition and properties of the entanglement entropy}
\label{sec:definition}
\vspace{-0.3cm}

\begin{wrapfigure}{l}{6cm}
\begin{center}
\resizebox{0.4\textwidth}{!} 
{\includegraphics{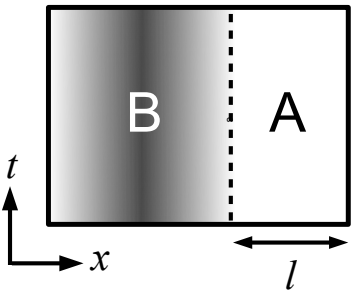}}
\vspace{-0.5cm}
\caption{
The complementary regions $A$ and $B$
separated by an imaginary boundary at $x=l$.
$y$ and $z$ axes are perpendicular to the plane.
Separation is purely an imaginary process and nothing
has to be done on the physical state.
The entanglement entropy measures quantum correlation
between two regions $A$ and $B$.
}
\label{fig:two_regions_AB}
\end{center}
\vspace{-0.5cm}
\end{wrapfigure}

The entanglement entropy of a pure state $|\Psi\rangle$
is defined as follows.
We divide the total system into subregion $A$ and its complement $B$.
See \Fig{fig:two_regions_AB}.
Let $l$ be the size of the system $A$ in the $x$ direction.
The density matrix of the system is
$
\rho = | \Psi \rangle \langle \Psi |.
$
Since we consider the pure state, the von Neumann entropy
of the system is clearly zero.
The reduced density matrix obtained by tracing out
the degrees of freedom in the region $B$,
\begin{equation}
\rho_A = \Tr_B \rho = \Tr_B | \Psi \rangle \langle \Psi |,
\end{equation}
describes the density matrix for an observer who can only access
to the subregion $A$.
Although we start off with a pure state with vanishing von Neumann entropy,
the state corresponding to the reduced density matrix is generally a mixed state.
$\rho_A$ contains the information on the quantum degrees of
freedom traced out.
The entanglement entropy is defined as the von Neumann entropy
of the reduced density matrix,
\begin{equation}\label{eq:entanglementE}
S_A = - \Tr \rho_A \ln \rho_A.
\end{equation}

\begin{wrapfigure}{r}{5cm}
\begin{center}
\resizebox{0.4\textwidth}{!}{\includegraphics{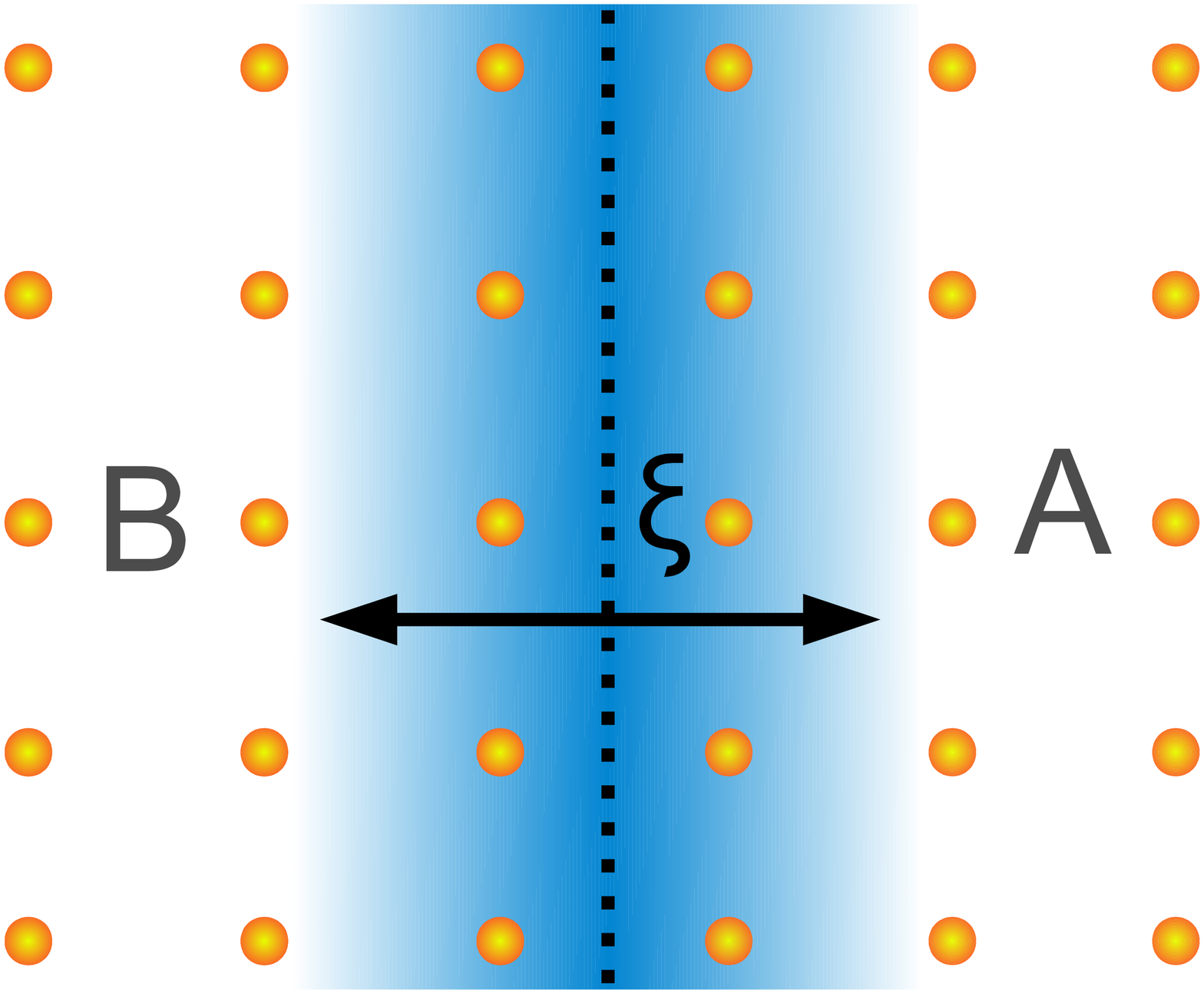}}
\vspace{-0.5cm}
\caption{
Schematic figure of spin system with a finite correlation length $\xi$.
Spin degrees of freedom in two regions $A$ and $B$ separated
more than the correlation length do not have quantum correlations,
and do not contribute to quantum entanglement.
}
\label{fig:correlation}
\end{center}
\vspace{-0.5cm}
\end{wrapfigure}
If the system has a finite mass gap (or a finite correlation length),
then the most of contributions to the entanglement between two regions
comes from the field degrees of freedom near the boundary
(see \Fig{fig:correlation}).
That is, those in $A$ and $B$ separated by more than the correlation length
have no quantum correlations and do not contribute to entanglement.
Accordingly, the entanglement entropy saturates above some critical length
in quantum field theories with mass gap
\cite{Wolf:2007dk}.
For instance, the entanglement entropy of the ground state
of the Ising chain model in the non critical regime
saturates at large $l$,
$S_A(l) \to c/3 \log \left( \xi/a \right)$.
Here $\xi$ is the correlation length of the system
and $a$ the lattice spacing, $c$ the central charge.

Although an analytic proof is still lacking,
numerical evidence of the existence of the mass gap
in Yang-Mills theory have been accumulated in 30 years,
and we expect that such a saturation can be seen in lattice QCD simulations.
Some other properties of the entanglement entropy can be found in
\cite{Nielsen:2000}.

\begin{wrapfigure}{l}{5.5cm}
\begin{center}
\resizebox{0.3\textwidth}{!} 
{\includegraphics{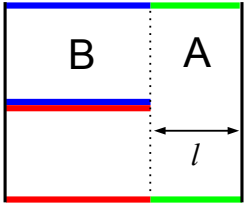}}
\caption{
Schematic picture for the system with two cuts
in $x-t$ plane.
In the region $A$ ($B$), the periodic boundary condition is imposed
with the period $2/T$ ($1/T$).
}
\label{fig:replica_trick_2}
\vspace{-0.5cm}
\end{center}
\end{wrapfigure}

\vspace{-0.3cm}
\section{Replica trick}
\label{sec:replica}

In order to evaluate the entanglement entropy,
we applied the replica trick.
The detail of the derivation is given in
\cite{Calabrese:2004eu}.
The point is that the entanglement entropy defined in
\Eq{eq:entanglementE} can be represented in the form,
$
S_A = - \lim_{n \to 1} \partial/\partial n
\ln \Tr_A \rho_A^n.
$

The trace of the $n$-th power of the reduced density matrix $\rho_A$
is given by the ratio of the partition functions,
\begin{equation}
\Tr \rho_A^n = Z(l,n)/Z^n.
\end{equation}
Here $Z(l,n)$ is the partition function of the system having
special topology, the $n$-sheeted Riemann surface.
The field variables in the region $A$ is periodically identified
with the interval $n/T$ ($T$ is the lattice extent in the temporal
direction, corresponding to temperature) while in the region $B$
the periodic boundary condition is imposed with the period $1/T$.
The case for $n=2$ is illustrated in \Fig{fig:replica_trick_2}.
$Z$ is the partition function with no cut.

The entanglement entropy is then given by
\begin{equation}
S_A(l) = - \lim_{n \to 1} \frac{\partial}{\partial n}
\ln \left( \frac{Z(l,n)}{Z^n} \right).
\end{equation}
The derivative of $S_A(l)$ with respect to $l$, which is
free of the ultraviolet divergence, can be expressed as follows;
\begin{equation}\label{eq:dSdl}
\frac{\partial S_A(l)}{dl}
= \frac{\partial}{\partial l} \left[ - \lim_{n \to 1} 
\frac{\partial}{\partial n} \ln \left( \frac{Z(l,n)}{Z^n}
\right) \right]
= \lim_{n \to 1}
\frac{\partial}{\partial l} \frac{\partial}{\partial n} F[l,n].
\end{equation}
That is, in order to calculate $\partial S_A/\partial l$,
we first evaluate the free energy of the system having $n$ cuts
with the length $l$ of the cut, then take the derivative
with respect to $n$ and $l$, and take the limit $n \to 1$.
Thus, the evaluation of the entanglement entropy is reduced
to calculate the free energy of the system with $n$ cuts.

\section{Lattice setup and observables}
\label{sec:lattice_setup}
\vspace{-0.3cm}

In numerical simulations, the derivative in \Eq{eq:dSdl}
have to be replaced by the finite difference,
and we estimate the derivative by
\begin{equation}\label{eq:dSdl_lat}
\lim_{n \to 1} \frac{\partial}{\partial l}
\frac{\partial}{\partial n} F[A,n]
 \to 
\frac{\partial}{\partial l}
\lim_{n \to 1} \left( F[l,n+1] - F[l,n] \right)
 \to  \frac{F[l+a,n=2] - F[l,n=2]}{a}.
\end{equation}
In the first line, the derivative with respect to $n$ is replaced
by the finite difference between the free energies for $n$ and $n+1$ cuts.
To go to the second line, we substitute $n=1$ to the free energies.
At this point, $\partial F[l,n=1]/\partial l$ drops out
since $F[l,n=1]$ does not depend on $l$.

The differences of free energies can be evaluated numerically
by introducing an `interpolating action' which interpolate
two actions corresponding to two free energies
\cite{Endrodi:2007dq,Fodor:2007sy},
$S_{\textrm{int}} = (1-\alpha) S_l[U] + \alpha S_{l+a}[U]$.
$S_l$ and $S_{l+a}$ represents the actions
corresponding to $F[l,n=2]$ and $F[l+a,n=2]$ in \Eq{eq:dSdl_lat}.
It is easy to show that
\begin{equation}\label{eq:alpha_integral}
  F[l+a,n=2] - F[l,n=2]
  = - \int^1_0 d\alpha \frac{\partial}{\partial \alpha} \ln Z(l,\alpha)
  = \int^1_0 d\alpha \left\langle S_{l+a}[\phi] - S_l[\phi] \right\rangle_{\alpha}.
\end{equation}
Here $\langle \cdot \rangle_{\alpha}$ refers to the Monte Carlo average
with the interpolating action $(1-\alpha) S_l[U] + \alpha S_{l+a}[U]$.
Therefore, the entanglement entropy can be evaluated numerically
by updating gauge configurations with the interpolating action
and calculating the action differences for various $\alpha$ and
perform a numerical integration over $\alpha$.
In order to evaluate the integral in \Eq{eq:alpha_integral},
we calculated the action differences from $\alpha=0$ to 1
by the step 0.1, and employed the Simpson's rule
to evaluate the integration numerically, which interpolates
neighboring points by a quadratic curve.

The lattice configurations are generated by
the heat-bath Monte Carlo technique with
the standard Wilson plaquette action.
In our simulations,
the first 5000 sweeps are discarded for thermalization,
and the measurement has been done every 100 sweeps.
The number of configurations for each $\beta$ and lattice
size is around 3000 to 8000.

\begin{wrapfigure}{r}{5.8cm}
\begin{center}
\resizebox{0.3\textwidth}{!} 
{\includegraphics{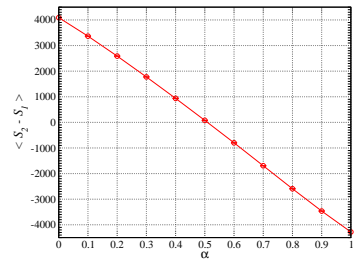}}
\vspace{-0.3cm}
\caption{
The difference $\langle S_{l=2} - S_{l=1} \rangle$
on $16^3 \times 32$ at $\beta=6.0$.
The integration from $\alpha=0$ to $\alpha=1$ gives
$\partial S_A/\partial l$ at $l=3a/2$,
the midpoint between $l=a$ and $l=2a$.
}
\label{fig:dS_1632B600L01}
\end{center}
\end{wrapfigure}

In \Fig{fig:dS_1632B600L01}, we plotted the action differences
$\langle S_{l=2} - S_{l=1} \rangle$
on $16^3 \times 32$ at $\beta=6.0$.
As is clear from the figure,
the line connecting the data points crosses zero at about $\alpha=0.5$.
Thus, most of the contribution cancels in integration over $\alpha$ from 0 to 1
though the absolute values of the differences are large at both end points.

\section{Simulation results}
\label{sec:results_dSdl}
\vspace{-0.3cm}

\noindent
{\bf Entanglement entropy}
\label{subsec:entanglement_entropy}

\begin{figure}
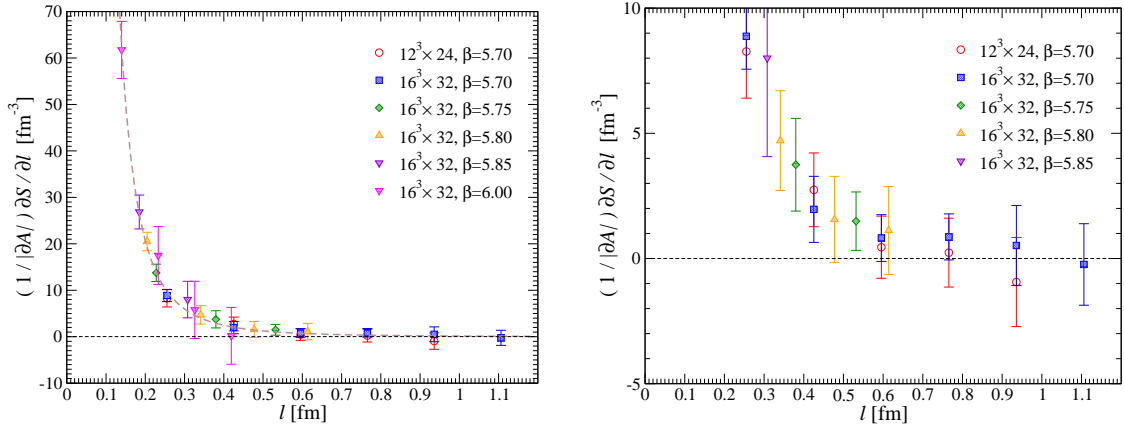

\begin{minipage}{0.47\hsize}\begin{center}
\resizebox{1.0\textwidth}{!}{\includegraphics
{dSdl_SU3}
}
\end{center}\end{minipage}
\hspace{0.03\hsize}
\begin{minipage}{0.47\hsize}\begin{center}
\resizebox{1.0\textwidth}{!}{\includegraphics
{dSdl_SU3_zoom}
}
\end{center}\end{minipage}
\caption{
The derivative $\frac{\partial S_A}{\partial l}$
of the entanglement entropy $S_A$ normalized by the area
$\partial A$ of the common boundary
with respect to the length of the region $A$.
The dashed curve is the fit of the data by the function
$c/l^{\alpha}$ with the fitted values $c=0.149(48), \alpha=3.06(20)$.
The right panel shows the zoom up of the left panel
to make near-zero region more visible.
}
\label{fig:dSdl_SU3}
\vspace{-0.5cm}
\end{figure}

The derivative of $S_A(l)$ with respect to $l$
is plotted in \Fig{fig:dSdl_SU3}.
$\partial S_A(l)/\partial l$ is normalized by the area
of the common boundary, $|\partial A|$.

We observe that data on $12^3\times 24$ and $16^3\times 32$
agree within statistical errors.
This implies that the derivative of the entanglement entropy
is proportional to the area of the boundary as is expected.

As is explained above, the entanglement entropy is closely
related to the correlation length $\xi$ of the Hamiltonian.
Thus, in small $l$ regions, the entanglement entropy is
expected to scale as $1/l^2$ from the dimensional analysis.
That is, $\partial S_A/\partial l$ behaves as $1/l^3$ at small $l$.
This behavior is exactly what the entanglement entropy
in conformal field theory in (3+1)-dimensional spacetime shows.
In order to confirm this, we fitted data with the function
$
\partial S_A/\partial l
= c \left( 1/l \right)^{\alpha},
$
and we obtain
$
c = c=0.149(48),
\alpha=3.06(20),
\chi^2/ndf = 0.192.
$
The fitted function is plotted in \Fig{fig:dSdl_SU3}
by dasshed curve.

The string feature of our result is that
the derivative of the entanglement entropy hits zero
at about $l^{\ast} = 0.6 \sim 0.7$ [fm] and vanishes above.
This means that the entanglement entropy does not increase
with increasing $l$ at large $l$ and the QCD vacuum has
a finite correlation length (or a finite mass gap).
We note that the critical temperature $T_c$ of
SU(3) pure Yang-Mills theory is estimated from
the behavior of the Polyakov loop susceptibility as 280 [MeV],
and $1/T_c \sim 0.714$ [fm]
\cite{Iwasaki:1996sn}.
This value and our result promise the identification,
$
l^{\ast} = 1/T_c.
$
In other words, the critical length of entanglement entropy
and (the inverse of) the critical temperature of
the deconfinement phase transition may be identified.

\noindent
{\bf Entropic $C$-function}
\label{subsec:c_function}

\begin{figure}
\begin{center}
\resizebox{0.4\textwidth}{!} 
{\includegraphics{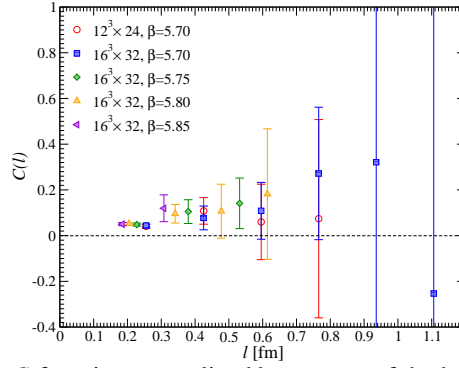}}
\vspace{-0.5cm}
\caption{
The entropic $C$-function normalized by an area of the boundary,
$C(l) = \frac{l^3}{|\partial A|} \frac{\partial S_A}{\partial l}$.
}
\label{fig:C_function_SU3}
\end{center}
\vspace{-0.5cm}
\end{figure}

The holographic analysis of the entanglement entropy
for confining backgrounds have revealed that there is a transition
from the connected solution to the disconnected solution
corresponding to $O(N_c^2)$ and $O(1)$ solutions, respectively
\cite{Klebanov:2008tg,Nishioka:2007zl}.
This predicts a jump in the entropic $C$-function,
$
C(l) = l^3/|\partial A| \partial S_A/\partial l,
$
at some critical length $l^{\ast}$, above which the $C$-function
vanishes.
This may be parallel to the confinement/deconfinement phase transition.

The numerical result for the entropic $C$-function
is shown in \Fig{fig:C_function_SU3}.
We observe that the $C$-function takes non-zero value
below $l^{\ast} \sim 0.6$ [fm].
Above 0.6 [fm], the numerical data suffer from huge statistical errors.
This is because $\partial S_A/\partial l$ is very small in this region
while its statistical errors do not so much depend on $l$.
Thus, the relative statistical error becomes quite large at large $l$,
and we cannot specify the critical length precisely.

\section{Summary and conclusion}
\label{sec:summary}
\vspace{-0.3cm}

We studied the entanglement entropy of the QCD vacuum
in SU(3) Yang-Mills theory using lattice Monte Carlo simulations.
The entanglement entropy is defined as the von Neumann entropy
of the reduced density matrix which is obtained by tracing out
the degrees of freedom in one of two complementary regions.
It measures the quantum correlation between subregions.
We find that the entanglement entropy scales as $1/l^2$
at small $l$ as in the conformal field theory.
The derivative of the entanglement entropy with respect to $l$
hits zero at about $l^{\ast} = 0.6 \sim 0.7$ [fm] and it vanishes
above this length.
It implies that the Yang-Mills theory has the mass gap
of the order of $1/l^{\ast}$.
This value is very close to the critical temperature
of the confinement/deconfinement phase transition.
The entropic $C$-function is suffered from large statistical errors
and we cannot specify the critical length precisely.
Simulations with the renormalization-group improved action
will improve this situation.

\noindent
{\bf Acknowledgements}

The simulation was performed on
NEC SX-8R at RCNP, Osaka University
and NEC SX-9 at CMC, Osaka University.
Y. N. is supported by Grant-in-Aid for JSPS Fellows
from Monbu-kagakusyo.
The work is partially supported by
Grant-in-Aid for Scientific Research by
Monbu-kagakusyo, No. 20340055.
We are gratefully acknowledge useful discussions 
on AdS/CFT correspondence with K. Nawa.

\vspace{-0.5cm}
\begin{appendix}

\section{Riemann sheets and gauge invariance}
\label{sec:gauge_invariance}
\vspace{-0.3cm}

\begin{figure}
\begin{minipage}{0.47\hsize}\begin{center}
\resizebox{0.8\textwidth}{!}{
\includegraphics{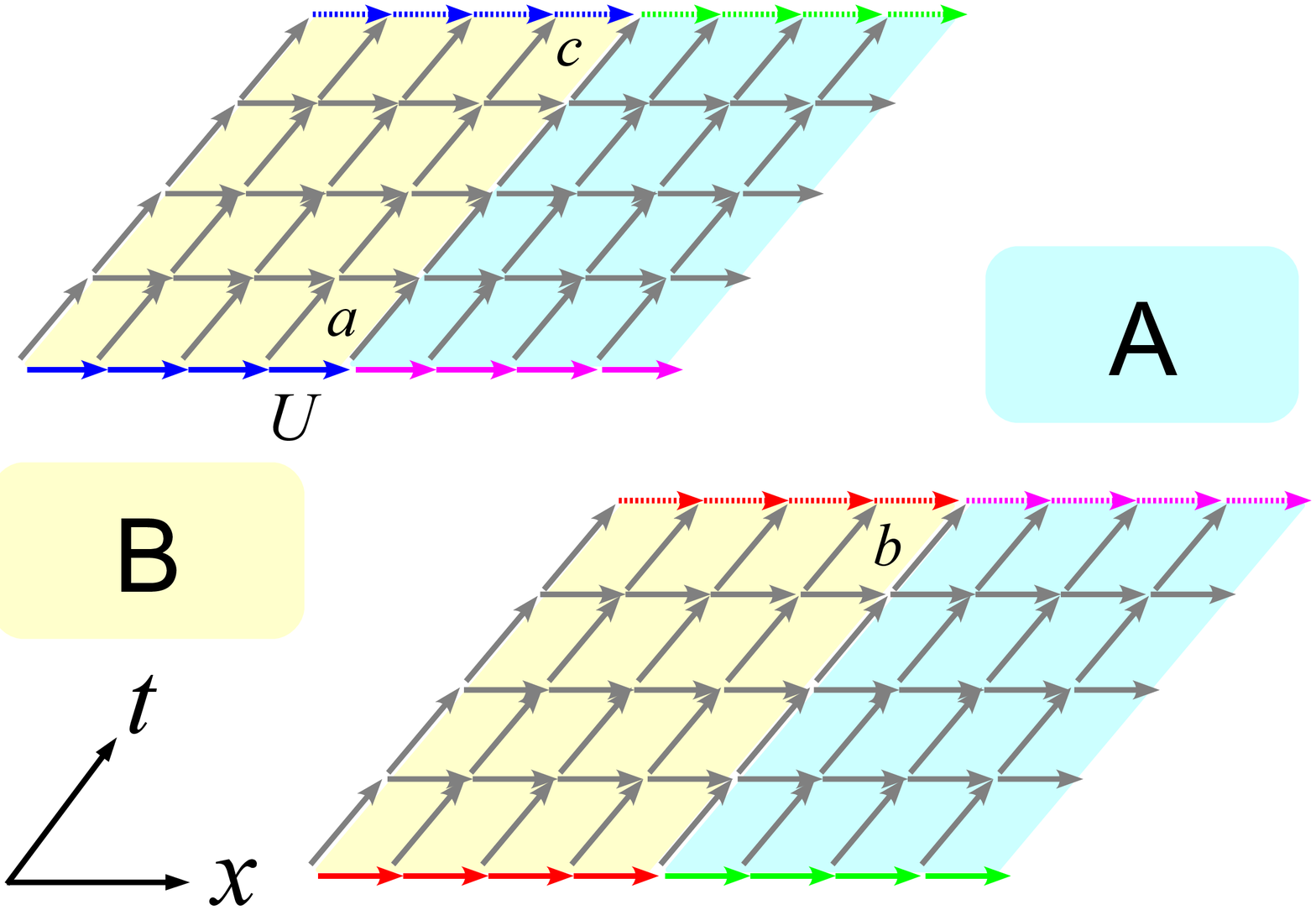}
}
\end{center}\end{minipage}
\hspace{0.03\hsize}
\begin{minipage}{0.47\hsize}\begin{center}
\resizebox{0.8\textwidth}{!}{
\includegraphics{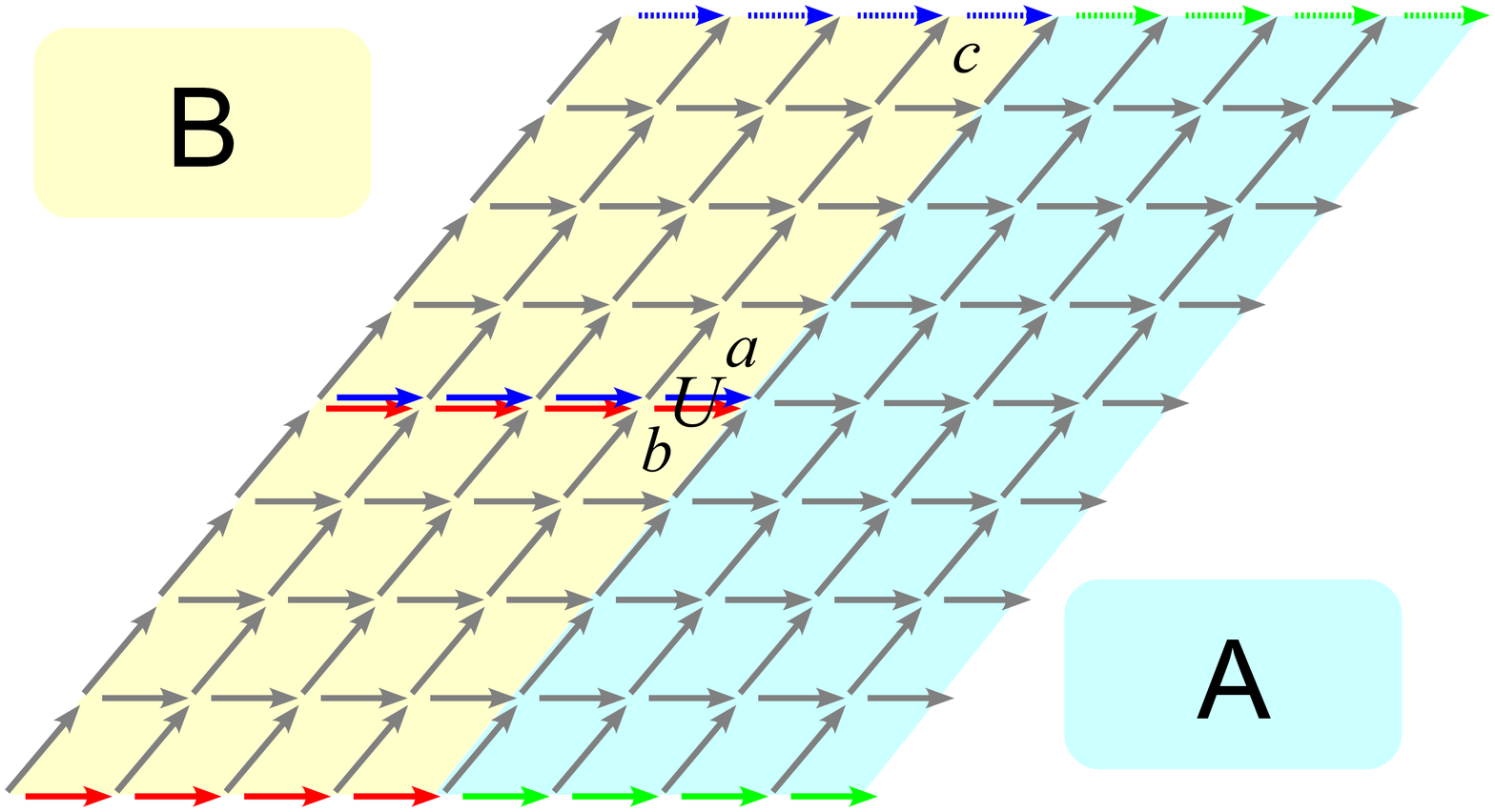}
}
\end{center}\end{minipage}
\caption{
(Left) Riemann surface structure of the lattice system with two cuts.
Links with same colors indicate periodically identified links.
(Right) Two Riemann sheets projected on a single plane.
In order to update the link variable denoted by $U$
without violating the gauge invariance,
we have to use the products of links (plaquettes) enclosing
the area $a$ and $c$, not $a$ and $b$.
}
\label{fig:Riemann_surface}
\vspace{-0.5cm}
\end{figure}

In lattice gauge theory, the Monte Carlo update
of link variables on $n$-sheeted Riemann surface
needs special care in order not to violate the gauge invariance.

Consider a lattice of the volume $8^3 \times 4$ and double the lattice
in the temporal direction to calculate the free energy
of the lattice with two cuts, see \Fig{fig:Riemann_surface}.
We denote links on two lattices as $\{U\}$ and $\{U'\}$.
The periodic boundary condition in the region $B$ is such that
(for simplicity, we suppress the Lorentz indices)
\vspace{-0.3cm}
\begin{itemize}
\item $U(\vec{x},t=0) = U(\vec{x},t=4)$
(red arrows in \Fig{fig:Riemann_surface})
\vspace{-0.3cm}
\item $U'(\vec{x},t=0) = U'(\vec{x},t=4)$
(blue arrows in \Fig{fig:Riemann_surface}),
\vspace{-0.3cm}
\end{itemize}
and that in the region $A$ is
\vspace{-0.3cm}
\begin{itemize}
\item $U(\vec{x},t=0) = U'(\vec{x},t=4)$
(green arrows in \Fig{fig:Riemann_surface})
\vspace{-0.3cm}
\item $U(\vec{x},t=4) = U'(\vec{x},t=0)$
(magenta arrows in \Fig{fig:Riemann_surface}).
\vspace{-0.3cm}
\end{itemize}
In order to update the link variable denoted by $U$
in \Fig{fig:Riemann_surface}, we need six plaquettes,
two of them are lying in $x-t$ plane.
Naively, we use the plaquettes denoted by $a$ and $b$ in the right panel
for two of six plaquettes to update the link $U$.
However, this leads to the violation of the gauge invariance
because the plaquettes $a$ and $b$ belongs to
\textit{different Riemann sheets}.
Therefore, the link $U$ lies in the upper Riemann sheet
in \Fig{fig:Riemann_surface} does not make a closed loop
with the staple $b$, meaning the loss of gauge invariance.
We have to use the plaquette denoted by $c$ to update the link $U$
otherwise the gauge invariance is not preserved.

\end{appendix}

\vspace{-0.3cm}


\end{document}